\documentclass[copyright,creativecommons]{eptcs}

\usepackage{todonotes}
\usepackage{listings}
\usepackage{comment}
\usepackage{float}
\usepackage{enumitem}
\usepackage{wrapfig}

\newfloat{listing}{tp}{lst}
\floatname{listing}{\bf Listing}
\floatstyle{plain}

\newcommand{\Listref}[1]{Listing~\ref{#1}}

\usetikzlibrary{shapes.geometric}

\newcommand{\tikzbox}[2]{\makebox(0,6)[r]{%
\tikz[remember picture,baseline]\node[very thick,red,anchor=base,rectangle,rounded corners,draw](#1){\phantom{#2}};%
\hspace*{-0.8ex}}}

\newcommand{\tikzarrow}[2]{\tikz[remember picture,overlay]\draw[very thick,red,->,solid] (#1) -- (#2);}

\begin{document}

\title{Verifying Parallel Loops with Separation Logic\thanks{This
  work is supported by the EU FP7 STREP project CARP (project nr. 287767).}}

\def\titlerunning{Verifying Parallel Loops with Separation Logic}

\author{
Stefan Blom
\qquad\qquad
Saeed Darabi
\qquad\qquad
Marieke Huisman
\institute{University of Twente\\
Enschede, The Netherlands}
\email{s.c.c.blom,s.darabi,m.huisman@utwente.nl}
}


\def\authorrunning{Blom, Darabi and Huisman}

\maketitle
\begin{abstract}
  This paper proposes a technique to specify and verify whether a loop
  can be parallelised. 
  Our approach can be used as an additional step in a
  parallelising compiler to verify user annotations about loop
  dependences. Essentially, our technique requires each loop iteration
  to be specified with the locations it will read and write. From the
  loop iteration specifications, the loop (in)dependences can be
  derived. Moreover, the loop iteration specifications also reveal
  where synchronisation is needed in the parallelised program.  The
  loop iteration specifications can be verified using permission-based
  separation logic.
\end{abstract}

\section{Introduction}

Parallelising compilers can detect loops that can be executed in
parallel. However, this detection is not perfect. Therefore
developers can typically also use a pragma to declare that a loop is parallel. Any loop annotated
with such a pragma will be assumed to be parallel by the compiler.

This paper addresses the problem of how to verify that loops that are
declared parallel by a developer can indeed safely be parallelised.
The solution is to add specifications to the program that when
verified guarantee that the program can be parallelised without
changing its meaning. Our specifications stem from permission-based
separation logic~\cite{BornatCOP05,Boyland03}, an extension of Hoare
logic. This has the advantage that we can easily combine the
specifications related to parallelisation with functional correctness
properties. 

We illustrate our approach on the PENCIL programming
language~\cite{DBLP:journals/corr/abs-1302-5586}.  This is a
high-level programming language to simplify using many-core
processors, such as GPUs, to accelerate computations.  It is currently
under development as a part of the CARP project\footnote{See
  \url{http://www.carpproject.eu/}.}.  However, our approach also
applies to other languages that use the concept of parallel loops,
such as OpenMP~\cite{dagum1998openmp}.  In order to simplify the
presentation in this paper, we limit ourselves to single loops.  At
the end of this paper, we will briefly discuss how to extend our
approach to nested loops.

Below, we first present some background information, and then we introduce the
specification language for parallel loops.  Next, we sketch how we can
implement automated verification of the specifications. Finally, we
conclude with future work.

\section{Background}

\paragraph{Parallel Hardware.} Modern hardware offers many different
ways of parallelising code. Most main processors nowadays are
multi-core. Additionally,
they often have a set of vector instructions that can operate on
small vectors instead of just a single value at once. 
Moreover, graphics processing units (GPUs) nowadays also can be
used for general-purpose programming.
Writing and tuning software for such accelerated
hardware can be a very time-consuming task.

\paragraph{The PENCIL Language.} The PENCIL programming language is
developed as a part of the CARP project. It is designed to be a
high-level programming language for accelerator programming, providing
support for efficient compilation. Its core is a subset of sequential
C, imposing strong limitations on pointer-arithmetic. In addition to traditional
C, it allows loops to be specified with two pragmas:
\emph{independent} and \emph{ivdep}, indicating that a loop can
be parallelised, because it is independent, or only contains forward
dependences, respectively.


\paragraph{Loop Dependences.}
Several kinds of loop dependences can be identified. There exists a
\emph{loop-carried dependence} from statement $S_{src}$ to statement
$S_{sink}$ in the body of a loop if there exist two iterations
\textit{i} and \textit{j} of that loop, such that:
\begin{itemize}[topsep=1pt,noitemsep]
\item[-] Iteration \textit{i} is before iteration \textit{j},
  \emph{i.e.}, $i < j$.
\item[-] Statements $S_{src}$ on iteration \textit{i} and $S_{sink}$ on iteration \textit{j} access the same memory location.
\item[-] At least one of these accesses is a write.
\end{itemize}
When $S_{src}$ syntactically appears before
$S_{sink}$ (or if they are the same) there is a
\textit{forward loop-carried dependence}, otherwise there is a \textit{backward loop-carried dependence}. The distance between two dependent iterations $i$ and $j$ is defined as the \textit{distance of dependence}.

\begin{wrapfigure}[10]{r}{.37\textwidth}
\vspace*{-3ex}
\begin{lstlisting}
 for(int i=1;i<=N;i++){
    #$S_1$#: a[i]#\tikzbox{src1}{a[i]}# = c[i] + 1;
    #$S_2$#: c[i] = a[i-1]#\tikzbox{dst1}{a[i-1]}# + 2;
 }
\end{lstlisting}
\tikzarrow{src1}{dst1}
\vspace*{-3ex}
\begin{lstlisting}
 for(int i=0;i<N;i++){
    #$S_1$#: a[i]#\tikzbox{dst2}{a[i]}# = c[i] + 1;
    #$S_2$#: c[i] = a[i+1]#\tikzbox{src2}{a[i+1]}# + 2;
 }
\end{lstlisting}
\vspace*{-1.8ex}
\tikzarrow{src2}{dst2}
\end{wrapfigure}


On the right, we show examples of first a forward and then a
backward loop carried dependence. In both cases there is a dependence between $S_1$ and
$S_2$. In the first loop, the read in $S_2$ reads the value
written in $S_1$ in the previous iteration of the loop. In the second loop,
the read in $S_2$ must be
done before the value is overwritten in $S_1$ during the next iteration.

The distinction between forward and backward dependences is
important. Independent parallel execution of a loop with dependences
is always unsafe, because it may change the result. However, a loop
with forward dependences can be parallelised by inserting an
appropriate synchronisation in the code, while loops with backward
dependences cannot be parallelised.


\paragraph{Separation Logic.}
Our approach to reason about loop (in)dependences uses
permission-based separation logic to specify which variables are read
and written by a loop iteration. Separation
logic~\cite{Reynolds02separationlogic} was originally developed as an
extension of Hoare logic to reason about pointer programs, as it
allows to reason explicitly about the heap. This makes it also suited
to reason modularly about concurrent programs~\cite{OHearn07}: two
threads that operate on disjoint parts of the heap do not interfere,
and thus can be verified in isolation. The basis of our work is a
separation logic for C~\cite{DBLP:conf/popl/TuchKN07}, but extended
with permissions~\cite{Boyland03}, to denote either the right to read from
or to write to a location. The set of permissions that a thread holds
are often known as its \emph{resources}. We write access permissions
as \lstinline+perm+\((e, \pi)\), where \(e\) is an expression denoting
a memory location and \(\pi \in (0,1]\) is a fraction, where any value permits reading and $1$ provides write permission.
The logic prevents the sum of permissions for a location over all threads to exceed $1$, which prevents data races.
In earlier work, we have shown that this logic is suitable to reason about kernel programs~\cite{BlomHM13}.


\section{A Specification Language for Loop Dependence}



The classical way to specify the effect of a loop is by means of an
invariant that has to hold before and after the execution of each
iteration in the loop.  Unfortunately, this offers no insight into
possible parallel execution of the loop. Instead we will consider
every iteration of the loop in isolation. To be able to handle
dependences, we specify restrictions on how the execution of the
statements for each iteration is scheduled.  In particular, each
iteration is specified by its own contract, \emph{i.e.}, its
\textit{iteration contract}. In the iteration contract, the
precondition specifies resources that a particular iteration requires
and the postcondition specifies the resources which are released after
the execution of the iteration. In other words, we treat
each iteration as a specified block~\cite{DBLP:conf/vstte/Hehner05}.

\Listref{Specification of independent loops}\ gives
an example of an \emph{independent loop},
specified by its iteration contract.  The contract requires that at
the start of iteration $i$, permission to write both \lstinline+c[i]+
and \lstinline+a[i]+ is available, as well as permission to read
\lstinline+b[i]+. 
The contract also ensures that these permissions are
returned at the end of iteration $i$. The iteration contract
implicitly requires that the separating conjunction of all iteration
preconditions holds before the first iteration of the loop, and that
the separating conjunction of all iteration postconditions holds after
the last iteration of the loop. In~\Listref{Specification of
  independent loops}, the loop iterates from $0$ to $N-1$, so the
contract implies that before the loop, permission to write the first
$N$ elements of both \lstinline+a+ and \lstinline+c+ must be
available, as well as permission to read the first $N$ elements of
\lstinline+b+. The same permissions are ensured to be available after
termination of the the loop.

\begin{listing}[t]
\begin{lstlisting} 
for(i=0;i<N;i+=1)
/*@ requires perm(a[i],1) ** perm(c[i],1) ** perm(b[i],1/2);
    ensures  perm(a[i],1) ** perm(c[i],1) ** perm(b[i],1/2); @*/
   { S1: a[i] = b[i] + 1;	
     S2: c[i] = a[i] + 2; }
\end{lstlisting}
\vspace*{-1em}\caption{Specification of an Independent Loop}\label{Specification of independent loops}
\end{listing}

To specify \emph{dependent loops}, in addition we need the ability to
specify what happens when the computations have to
synchronise due to a dependence.  During such a synchronisation, permissions should be
transferred from the iteration containing the source of a dependence
to the iteration containing the sink of that
dependence. 
To specify a
\emph{permission transfer} we introduce the \lstinline+send+ keyword:
\begin{lstlisting}
//@ send #$\phi$# to #$L$#, #$d$#;
\end{lstlisting}
This specifies that the permissions and properties expressed by the 
separation logic formula $\phi$ are transferred to the statement labelled $L$
in the iteration $i+d$, where $i$ is the current iteration and $d$ is the distance of dependence. 

\begin{listing}[tp]
\begin{lstlisting}
for(int i=1;i<=N;i++)  
/*@ requires i==1 ==> perm(a[i-1],1/2);
    requires perm(c[i],1) ** perm(a[i],1);
    ensures  perm(c[i],1) ** perm(a[i],1/2) ** perm(a[i-1],1/2);
    ensures  i==N ==> perm(a[i],1/2); @*/
{
    S1: a[i] = c[i]*CONST +a[i]*(1-CONST);	
    //@ send perm(a[i],1/2) to S2,1;
    #//# if (i>1) receive perm(a[i-1],1/2);
    S2: c[i] = min(a[i],a[i-1]); 
}
\end{lstlisting}
\caption{Specification of a Forward Loop-Carried Dependence}
\label{forward specified}
\end{listing}

\begin{listing}[tp]
\begin{lstlisting}
for( i=0;i<N;i++)
/*@ requires i==0 ==> perm(a[i],1/2);
    requires perm(c[i],1) ** perm(a[i],1/2) ** perm(a[i+1],1/2);
    ensures  perm(c[i],1) ** perm(a[i],1);
    ensures  i==N-1 ==> perm(a[i+1],1/2); @*/
{
   #//# if (i>0) receive perm(a[i],1/2);
   S1: a[i] = c[i]*CONST + a[i]*(1-CONST);					
   S2: c[i] = min(a[i+1],a[i]);		
   //@ send perm(a[i+1],1/2) to S1,1;  
}
\end{lstlisting}
\caption{Specification of a Backward Loop-Carried Dependence}
\label{backward specified}
\end{listing}

Below, we will give two examples that illustrate how loops are
specified with \lstinline+send+ clauses. The \lstinline+send+ clause
alone completely specifies both how permissions are provided and used
by the iterations. However, for readability, we also mark the place
where the permission are used with a corresponding \lstinline+receive+
statement as a comment.
\Listref{forward specified} gives a specified program with
a forward dependence, similar to our earlier example, 
while \Listref{backward specified}
gives an example of a program with a backward dependence.

We discuss the annotations of the first program in some detail. Each iteration $i$
starts with write permission on \lstinline+a[i]+ and \lstinline+c[i]+.
The first statement is a write to \lstinline+a[i]+, which needs write permission.
The second statement reads \lstinline+a[i-1]+, which is not allowed unless
read permission is available. For the first iteration, this read permission is available.
For all subsequent iterations, permission must be transferred.
Hence a \lstinline+send+ annotation is specified after the first assignment
that transfers a read permission on \lstinline+a[i]+ to the next
iteration (and in addition, keeps a read permission itself). The
postcondition of the iteration contract reflects this: it ensures
that the original permission on \lstinline+c[i]+ is released,
as well as the read permission on \lstinline+a[i]+,
which was not sent, and also the read permission on \lstinline+a[i-1]+,
which was received. Finally, since the last iteration cannot transfer
a read permission on \lstinline+a[i]+, the iteration contract's
postcondition also specifies that the last iteration returns this
non-transferred read permission on \lstinline+a[i]+.

The specifications in both listings are valid. Hence every execution
order of the loop bodies that respects the order implied by the
\lstinline+send+ annotations yields the same result as sequential
execution. In the case of the forward dependence example, this can be
achieved by adding appropriate synchronisation in the parallelised
code. All parallel iterations should synchronise each
\lstinline+send+ annotation with the location of the specified label to ensure proper permission
transfer.
For the backward dependence example, only sequential execution respects the ordering.

\section{Verifying Dependence Annotations}

To verify an iteration contract, we encode it as a standard method
contract that can be verified using the VerCors tool set~\cite{FM2014vercors}.
Suppose we have a loop specified with an iteration contract as below:
\begin{lstlisting}
  #$S_{\rm pre}$#;
  for(int i=0;i<N;i++)
  /*@ requires pre(i);
      ensures  post(i); @*/
  {   #$S$#; }
  #$S_{\rm post}$#;
\end{lstlisting}
To prove that this program respects its annotations, the following
proof obligations have to be discharged:
\begin{itemize}[topsep=1pt,noitemsep]
\item[-] after $S_{\rm pre}$, the separating conjunction of all of the
  iteration preconditions holds;
\item[-] the loop body $S$ respects the iteration contract; and
\item[-] the statement $S_{\rm post}$ can be proven correct, assuming that
the separating conjunction of the postconditions holds.
\end{itemize}

To generate these proof obligations, we encode the original program by
generating several annotated procedures by the following steps:
\begin{enumerate}[topsep=2pt,noitemsep]
\item We replace every loop in the program with a call to a procedure \lstinline+loop_main+,
whose arguments are the free variables occurring in the loop.
The contract of this procedure requires the separating conjunction of all preconditions and ensures
the separating conjunction of all postconditions. After this replacement, we can verify the
program with existing tools to discharge the first and the last proof
obligations.
\item 
To discharge the remaining proof obligation, we generate a procedure \lstinline+loop_body+,
whose arguments are the loop variable $i$
plus the same arguments as \lstinline+loop_main+.
The contract of this procedure is the iteration contract of the loop body,
preceded by a requirement that states that the value of the iteration variable is within
the bounds of the loop.
\end{enumerate}
The result of this encoding is as follows:

\begin{lstlisting}
void block(){
    #$S_{\rm pre}$#;
    loop_main(N,#${\sf free}(S)$#);
    #$S_{\rm post}$#;
}

/*@ requires (\forall* int i;0<=i && i<N; pre(i));
    ensures  (\forall* int i;0<=i && i<N; post(i)); @*/
loop_main(int N,#${\sf free}(S)$#));

/*@ requires (0<=i && i<N) ** pre(i);
    ensures  post(i); @*/
loop_body(int i,int N,#${\sf free}(S)$#)){ #$S$#; }
\end{lstlisting}


Verification of the \lstinline+send+ instruction
is done by replacing the \lstinline+send+ annotation with a procedure
call \lstinline+send_phi(i);+ and by inserting a procedure call \lstinline+recv_phi(i);+ at the location of the label $L$.
The contracts of these methods encode the transfer of the resources specified by $\phi(i)$
from the sending iteration to the receiving iteration, subject to two conditions:
\begin{enumerate}[topsep=2pt,noitemsep]
\item Permissions can only be transferred to future iterations ($d>0$).
\item Transfer only happens if both the sending and the receiving iterations exist.
\end{enumerate}
The existence of iteration $i$ is expressed by the predicate \lstinline+is_iteration(i)+,
whose definition is derived from the loop bounds. For example, the loop \lstinline?for(int i=0;i<N;i++)?
gives rise to
\begin{lstlisting}
  boolean is_iteration(int i){return 0 <= i && i < N;}
\end{lstlisting}
Using this notation the generated (abstract) methods and contracts are:
\begin{lstlisting}
/*@ requires is_iteration(#$i+d$#) ==> #$\phi(i)$#;
 @*/ 
void send_phi(int i);

/*@ ensures  is_iteration(#$i-d$#) ==> #$\phi(i-d)$#; 
@*/ 
void recv_phi(int i);
\end{lstlisting}
Note that instead of a constant $d$, we may use any invertible function $d(i)$.


\section{Conclusion and Future Work}

This paper sketches how to verify parallel loops, even in
the presence of dependences from one loop iteration to the next. The idea is to
specify each iteration of a loop with its own iteration contract and
to use the \lstinline+send+ annotation to transfer permission between iteration if needed.
We conjecture that if verification of a loop is possible without using
\lstinline+send+ then it is correct to tag the loop as
\lstinline+independent+, \emph{i.e.}, 
an iteration never reads a location that was written by a different iteration.
Moreover, if \lstinline+send+ is used with
labels occurring after the statement then it is correct to use PENCIL's
\lstinline+ivdep+ tag to indicate parallelisability.

The method described is modular in the sense that it allows us to treat
any parallel loop as a statement, thus nested loops can be dealt with simply by
giving them their own iteration contract. Alternatively one iteration contract
can be used for several nested loops.


It is future work to provide a formal proof for our conjecture, as
well as to develop fully automated tool support for discharging the
proof obligations. We also plan to link our PENCIL specifications with
our kernel logic~\cite{BlomHM13} and to define compilation of PENCIL
specifications. 

Another possible direction for future work is to extend our approach to reason about the correctness of OpenMP~\cite{dagum1998openmp} pragmas in parallel C programs.
From the point of view of verification, many concepts in OpenMP and PENCIL are the same.
For example, the \textbf{simd} pragma in OpenMP is used in the same way as PENCIL uses \lstinline+ivdep+.
In general, our method can be applied for verification of any high-level
parallel programming language which uses compiler directives for parallelisation. 

Finally, we will also investigate how the iteration contracts for the verifier and parallelisation
pragmas for the compiler can support each
other.  
We believe this support can work in both ways.
First of all, the parallelising compiler can use verified
annotations to know about dependences without analysing the code
itself. In particular, the PENCIL language has a feature, called {\em
  function summaries}, that allows the programmer to tell the compiler
which memory locations are written and/or read by a function by
writing a fake function that assigns to the writable locations and
reads from the readable locations.  Such summaries are easily
extracted from specifications, and thus in this way specifications can
help to produce
better code.  Conversely, if the compiler performs
an analysis then it could emit its findings as a specification
template for the code, from which a complete specification can be
derived.

\bibliographystyle{eptcs}
\bibliography{vercors,extra}
\end{document}